# Rethinking Issue Resolution for AI/ML Systems

Ahmed Adnan*
*East West University*
Dhaka, Bangladesh
bsse1131@iit.du.ac.bd

Mushfiqur Rahman*
*BUBT*
Dhaka, Bangladesh
mushfiqur.rahman@bubt.edu.bd

Antu Saha
*William & Mary*
Williamsburg, VA, USA
asaha02@wm.edu

Oscar Chaparro
*William & Mary*
Williamsburg, VA, USA
oscarch@wm.edu

***Abstract*—We advocate for AI/ML issue resolution frameworks tailored to maintenance workflows and the nature of modern AI/ML systems. Existing issue resolution frameworks largely emerged for traditional software maintenance practices and do not explicitly account for characteristics common in AI/ML systems, such as stochastic behavior, experimentation-driven workflows, and heterogeneous artifacts beyond source code. To identify the unique characteristics of issue resolution in AI/ML systems and motivate the need for tailored frameworks, we conducted a qualitative study of issue resolution workflows documented in 100 issue reports and pull requests across four widely used AI/ML systems: TensorFlow, scikit-learn, MLflow, and AutoGPT. Our findings suggest that issue resolution in AI/ML systems involves: recurring AI/ML-related activities that span multiple resolution stages; iterative experimentation and adaptive verification; and coordinated changes across artifacts such as datasets, prompts, and model configurations. We also observed challenges related to reproducibility, nondeterministic behavior, and artifact coordination. Building on these findings, we present a vision for AI/ML issue resolution frameworks and discuss research directions and tooling support needed to realize this vision.**

## I. The Vision

We advocate for issue resolution frameworks tailored for AI/ML systems. The increasing adoption of AI/ML development libraries and AI-integrated systems requires extending traditional issue resolution frameworks with AI/ML workflows that explicitly account for data, models, stochastic behavior, experimentation, and heterogeneous software artifacts. Existing issue resolution frameworks largely emerged from maintenance practices in traditional software systems, where failures are often more reproducible, debugging centers primarily around source code, and fixes are commonly validated through deterministic tests, code review, and static analysis [1]–[3]. However, modern AI/ML systems increasingly combine traditional software components with AI/ML artifacts such as datasets, prompts, models, hyperparameters, and training pipelines [4], [5]. Consequently, issue resolution in these systems may require developers to monitor model behavior, retrain models, tune hyperparameters, modify prompts or datasets, manage infrastructure configurations, and validate fixes using repeated experimentation, statistical reasoning, or qualitative evaluation procedures rather than relying solely on binary pass/fail judgments [6]–[9].

Within this vision, future issue resolution frameworks should treat AI/ML artifacts as first-class entities in maintenance workflows rather than as secondary components attached to source code. Such frameworks should support iterative and cross-stage issue resolution activities related to AI/ML components, reproducibility management for nondeterministic behaviors, performance-aware verification, and traceability across heterogeneous artifacts. For example, resolving a failure in a GenAI-based system may require developers to inspect prompts, datasets, model configurations, inference pipelines, and infrastructure environments while repeatedly evaluating model behavior under different execution conditions. We therefore envision AI/ML-adaptive issue resolution frameworks as extensions of existing frameworks that integrate support for experimentation, reproducibility, and heterogeneous artifact management into broader maintenance workflows.

To illustrate the unique characteristics of issue resolution in AI/ML systems and motivate the need for adapted frameworks, we conducted a preliminary case study. We qualitatively analyzed 100 issue reports and associated pull requests across four widely used AI/ML libraries/systems: TensorFlow [10], scikit-learn [11], MLflow [12], and AutoGPT [13]. Following an iterative open coding methodology [14], we investigated the resolution activities developers perform, the challenges they encounter, the mitigation strategies they adopt, and the artifacts they modify while resolving issues in AI/ML systems.

Our study reveals that issue resolution in AI/ML systems involves recurring AI/ML-related activities in addition to traditional issue resolution stages, heterogeneous artifacts beyond source code, and adaptive and non-deterministic validation and mitigation strategies that are not explicitly modeled in traditional frameworks. In particular, we observed that activities such as model performance monitoring and parameter tuning frequently span multiple issue resolution stages, suggesting that issue resolution in AI/ML systems is often more iterative and tightly coupled with experimentation workflows. Building on these findings, this paper presents our vision for AI/ML-specific issue resolution frameworks and discusses the maintenance workflows, tooling support, and research directions needed to realize this vision.

## II. Related Work

Prior work in software engineering for AI/ML systems (SE4ML) has shown that AI/ML systems differ from traditional software in their development workflows, architectures, and maintenance requirements [4], [15], [16]. They are data-driven, rely on complex training and deployment pipelines, and require continuous monitoring, retraining, and experimentation throughout the lifecycle [4], [15]. Research on MLOps and ML system maintenance identifies challenges in reproducibility, traceability, scalability, documentation, infrastructure management, and coordinating heterogeneous artifacts such as datasets, models, configurations, and pipelines [16]–[19]. Similarly, Jiang *et al.*'s study of pre-trained model reuse on Hugging Face [20] reports challenges with reproducibility, provenance, and nondeterministic behavior. The study focuses on *model selection and reuse* rather than on *issue resolution workflows* in AI/ML systems, as we do.

*Equal contribution

Prior work has also shown that AI/ML issues often stem from data quality, training configurations, model behavior, infrastructure dependencies, and non-deterministic execution rather than only production code [6], [7], [9]. Existing studies have proposed AI/ML fault taxonomies [9], analyzed the root causes and resolution of ML bugs [6], [7], [21], and investigated code-centric fault localization challenges [8]. For example, Humbatova *et al.* [9] found that deep learning faults frequently involve data, model configuration, and training processes, while Nguyen *et al.* [8] highlighted limited observability, non-determinism, and immature tooling as key barriers to bug localization.

Despite these advances, prior work focuses on broad AI/ML development practices, maintenance challenges, bug characteristics, and debugging techniques rather than on issue resolution worflows for AI/ML systems. Moreover, existing issue resolution frameworks in software engineering (SE) [1], [2], [22]–[24] have emerged for traditional systems. These frameworks model issue resolution as a sequence of stages such as analysis, reproduction, implementation, and verification, assuming deterministic behavior, code-centric debugging, and binary test outcomes [1], [3]. While prior SE4ML work broadly calls for adapting SE practices to AI/ML systems [4], [15], no prior work has specifically studied how issue resolution workflows should evolve. Our work addresses this gap by presenting a vision for AI/ML issue resolution frameworks grounded in preliminary evidence from four real-world AI/ML systems.

## III. Issue Resolution in AI/ML systems

We conducted a preliminary qualitative analysis of real-world AI/ML issue reports and associated artifacts to identify the unique aspects of issue resolution in AI/ML systems and the need for tailored frameworks currently missing in the field.

### A. Study Methodology

We briefly describe the study methodology, which is based on our prior qualitative study of issue report threads to identify issue resolution workflows in traditional systems [1]—a detailed description is available in our replication package [25].

*1) Project Selection:* We focused on two representative types of AI/ML systems: (i) frameworks/libraries to build AI/ML components, and (ii) software systems that integrate AI/ML models (including generative AI models) to implement various functionalities. Using GitHub's API [26], we searched for relevant GitHub repositories based on the following criteria: (i) open-source, (ii) popular (>100 stars), (iii) actively maintained (*i.e.*, with at least one commit/month over the past 6 months with a minimum of 5 active contributors), and (iv) had a large issue collection (>1000 closed issues). From an initial pool of 15 candidates, we selected four representative systems: TensorFlow [10] and scikit-learn [11] (as AI/ML frameworks/libraries), and MLflow [12] and AutoGPT [13] (as MLOps and GenAI-integrated systems, respectively).

*2) Issue Collection:* We analyzed closed issue reports from 2020 to 2025, encompassing bug reports, enhancement requests, new feature implementations, *etc.* From each project, 25 issues were randomly sampled, yielding a total of 100 issues. We limited our sample to 25 issues per project to keep the qualitative coding effort manageable, as each issue and its associated pull requests required detailed multi-coder annotation, inspection of different commits and artifacts, and reconciliation. Closed issues allow us to understand the full process as they provide end-to-end evidence of the whole resolution process, including problem analysis, solution design, implementation, and verification.

*3) Issue Coding Procedure:* Inspired by our prior work [1], we employed an iterative multi-coder open coding methodology [27] to annotate textual snippets (*e.g.*, phrases, sentences, and paragraphs) present in the discussion threads and various contributor actions (*e.g.*, creating commits, merging PRs, *etc.*) in the sampled issues and associated pull requests. The first two authors, both with an M.S. in SE and $\sim$2 years of industry experience in software maintenance and issue management, independently analyzed and assigned fine-grained *codes* to text snippets from the 100 issues in batches of 10 using the Hypothesis annotation tool [28]. The third and four authors, a Ph.D candidate and a professor, guided and reviewed the annotation process.

Starting from our prior study's codebook [1], we iteratively expanded the code catalog to capture AI/ML-related activities (*e.g.*, model retraining and evaluation), challenges, mitigation strategies, and associated traditional resolution stages (*e.g.*, solution design, implementation, *etc.*). The coders annotated actionable issue resolution activities as well as discussion content describing developer difficulties or mitigation approaches. After each batch, the coders conducted reconciliation sessions to resolve disagreements, refine coding guidelines, and retrospectively apply codebook/coding updates to maintain consistency. Instead of calculating inter-coder agreement metrics, reliability was established throughout the coding process through critical, consensus-driven discussions and iterative refinement, consistent with qualitative (inductive) open-coding practices [14].

*4) Coding Results:* We annotated 988 snippets across 100 issues, including 947 traditional issue-resolution activities and 41 AI/ML-specific unique activities. These issues were categorized as AI/ML (64), hybrid (18), or non-AI/ML (18) based on whether fixes/implementations were applied in AI/ML, traditional, or hybrid features or components.

*5) Thematic Analysis:* From thematic analysis of fine-grained codes, we synthesized higher-level categories representing issue resolution activities, challenges, and mitigation strategies. For instance, activities such as *'model accuracy evaluation'* and *'training loss tracking'* were grouped into a broader activity: MODEL PERFORMANCE MONITORING. Similarly, challenges such as *'version conflict', 'missing dependency', and*

TABLE I: AI/ML-related issue resolution activities (# of issues in parentheses)

| AI/ML Activities | Associated Stages | Projects |
|---|---|---|
| MODEL PERFORMANCE MONITORING: Continuously assessing models' predictive performance or the impact of bug fixes or changes on performance or resource utilization (22 issues) | SOLUTION DESIGN (11), IMPLEMENTATION (5), REPRODUCTION (4), ISSUE ANALYSIS (6), VERIFICATION (2) | Tensorflow (8), Scikit-learn (9), MLflow (4), AutoGPT (1) |
| PARAMETER TUNING AND TRAINING: Adjusting parameters and retraining models to improve performance (9 issues) | IMPLEMENTATION (7), SOLUTION DESIGN (2), VERIFICATION (1) | Tensorflow (3), Scikit-learn (3), MLflow (2), AutoGPT (1) |
| DATA MODIFICATION: Improve implementation and performance by updating inconsistent training data instances (6 issues) | IMPLEMENTATION (4), SOLUTION DESIGN (2) | Tensorflow (1), Scikit-learn (2), MLflow (2), AutoGPT (1) |
| MODEL FUNCTIONALITY ANALYSIS: Inspecting model architectures to understand outputs and pipeline interactions (4 issues) | SOLUTION DESIGN (3), IMPLEMENTATION (2) | Tensorflow (1), Scikit-learn (1), AutoGPT (2) |

*'library mismatch'* were grouped into the broader challenge *'dependencies & package version mismatching'*. This process resulted in 10 actionable activity categories (6 traditional stages and 4 AI/ML-specific activities) and 8 challenges.

We grounded our analysis in an established issue resolution taxonomy [1] and considered only the actionable stages directly involved in the resolution process (*e.g.*, solution design, implementation, etc.) from this taxonomy to annotate our issues. During annotation, we identified several AI/ML-specific activities that did not align with these existing stages (*e.g.*, activities related to data modification, model performance monitoring, *etc.*). To identify challenges, we analyzed and coded recurring difficulties faced by contributors across issue discussion threads.

### B. Study Findings

Our preliminary study provides evidence that issue resolution in AI/ML systems extends beyond traditional deterministic and code-centric workflows. Across 100 issues, developers repeatedly reasoned about model behavior, datasets, prompts, training configurations, and execution environments while navigating challenges related to nondeterminism, reproducibility, and heterogeneous artifacts through iterative experimentation and adaptive validation. These characteristics introduce maintenance concerns that existing issue resolution frameworks do not explicitly model, motivating the need for AI/ML-tailored frameworks.

*1) Finding 1 – AI/ML Issue Resolution Introduces Additional AI/ML-related Activities:* Developers resolving issues in AI/ML systems perform traditional issue resolution stages, identified in our prior work [1] with varying frequencies (REPRODUCTION: in 42% of the issues, ISSUE ANALYSIS: 65%, SOLUTION DESIGN: 62%, IMPLEMENTATION: 54%, CODE REVIEW: 34%, and VERIFICATION: 53%). However, beyond these stages, we observed four recurring AI/ML-related activities that required developers to reason about AI/ML-specific artifacts and behaviors: MODEL PERFORMANCE MONITORING (27%), PARAMETER TUNING AND TRAINING (11%), DATA MODIFICATION (7%), and MODEL FUNCTIONALITY ANALYSIS (5%) (see Table I). MODEL PERFORMANCE MONITORING and PARAMETER TUNING AND TRAINING occurred more frequently in TensorFlow and scikit-learn, likely because these repositories focus on model development and training, where monitoring model behavior and tuning training parameters are central maintenance activities. We also observed that developers use MODEL PERFORMANCE MONITORING throughout issue analysis, implementation, and verification to evaluate predictive performance and resource utilization. PARAMETER TUNING AND TRAINING stabilizes model behavior, DATA MODIFICATION corrects training data, and MODEL FUNCTIONALITY ANALYSIS helps diagnose unexpected predictions.

Fig. 1: Mapping between AI/ML-related issue resolution activities, traditional stages, and challenges

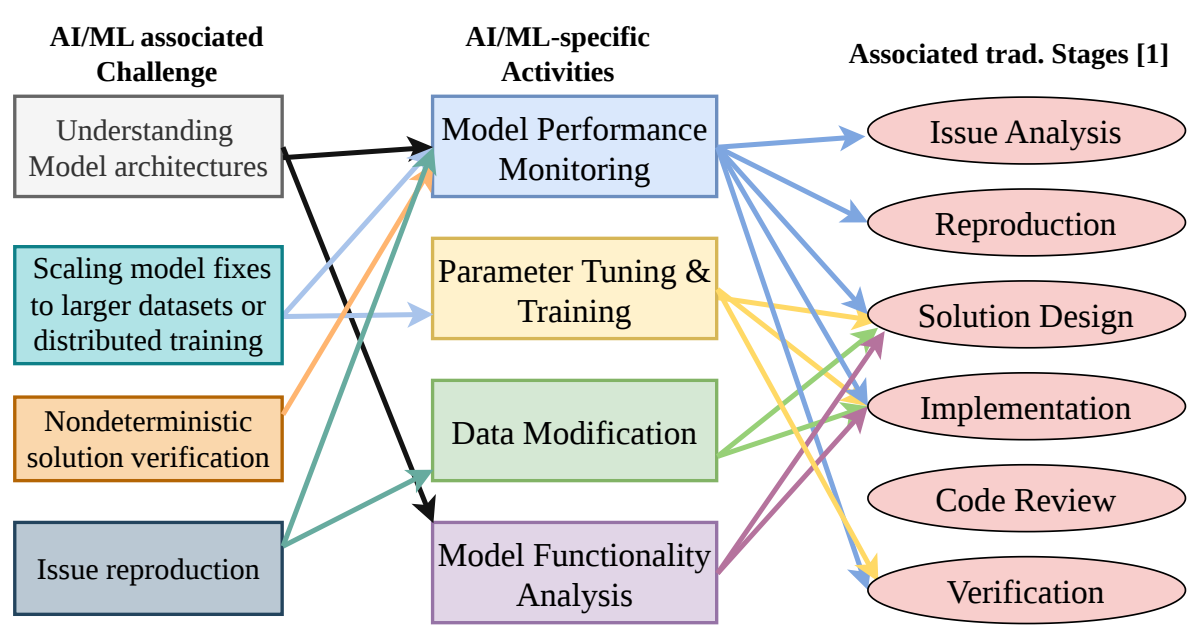

*2) Finding 2 – AI/ML-related Activities Often Span Multiple Resolution Stages:* Our study further suggests that these AI/ML-related activities are often cross-cutting and recur across multiple traditional issue resolution stages rather than appearing as isolated phases (see Table I and fig. 1).

For example, developers perform MODEL PERFORMANCE MONITORING during ISSUE ANALYSIS to diagnose unexpected behavior, during SOLUTION DESIGN and IMPLEMENTATION to evaluate candidate fixes, and during VERIFICATION to assess post-fix predictive performance and stability. Similarly, PARAMETER TUNING AND TRAINING frequently reappears as developers repeatedly retrain and reevaluate models after modifying datasets, prompts, configurations, or implementation logic. This suggests that instead of following a strictly sequential workflow, developers often refine solutions iteratively through repeated experimentation, monitoring, retraining, and reevaluation.

*3) Finding 3 – AI/ML Systems Introduce Additional Resolution Challenges:* Our study suggests that issue resolution in AI/ML systems include challenges related to stochastic behavior, evolving datasets, complex model architectures, and distributed execution environments (see Table II). One recurring challenge involves nondeterministic behavior during REPRODUCTION and VERIFICATION. We observed cases where reproducing the same issue across environments yields different outcomes because model behavior depended on factors such as random initialization, data sampling, hardware configurations, dependency versions, and runtime environments. For example, developers in TensorFlow's issue #46168 [29] struggled to determine whether a patch genuinely improved model accuracy

TABLE II: Issue resolution challenges and mitigation approaches in AI/ML systems (# of issues in parentheses)

| Challenge (Freq.) | Mitigation Approaches (Freq.) |
|---|---|
| Scaling model fixes to larger datasets or distributed training (9) | Statistical tests and repeated executions (6) [31]; distributed validation (2) [32]; tracking memory consumption using PyTorch (1) [33] |
| Understanding model architectures (6) | Structured code documentation (5) [30]; architecture inspection and pipeline analysis (1) [34] |
| Nondeterministic solution verification (4) | Statistical test validation (3) [29]; repeated runs with random seeds (1) [35] |
| Issue reproduction (4) | Containerized environments with version control (3) [36]; fixing data inconsistencies (1) [37] |

because repeated reproductions produced inconsistent results.

We also observed challenges associated with understanding model architectures and scaling fixes. Developers sometimes struggle to reason about complex model pipelines and internal behaviors when diagnosing issues (*e.g.*, MLFlow's issue #12790 [30]). Similarly, fixes that appear effective in small-scale experiments occasionally degrade predictive performance when executed on larger datasets or distributed environments (*e.g.*, MLFlow's issue #5653 [31]). Although these challenges appear in a subset of issues rather than universally across all, they repeatedly emerge across multiple systems and often intersect with multiple activities and stages (see Figure 1).

*4) Finding 4 – Developers Adopt Adaptive Validation and Mitigation Strategies:* To address AI/ML-specific challenges, developers adopt mitigation strategies tailored to the stochastic and experimentation-driven nature of AI/ML systems (see Table II). Unlike traditional issue resolution, where deterministic validation and fixed test suites are often sufficient, developers in AI/ML systems frequently rely on adaptive and iterative validation practices (see Table II).

For example, developers perform repeated executions with different random seeds, conduct statistical comparisons across runs, and continuously monitor predictive performance before merging fixes. In several issues, developers also validate fixes under distributed or production-like environments to ensure stable behavior at scale. Developers additionally adopt strategies to improve reproducibility and coordination across environments, including containerized environments, dependency management, and detailed documentation of datasets, pipelines, and model configurations. Developers therefore rely on repeated experimentation, statistical evaluation, and qualitative inspection rather than deterministic pass/fail testing.

*5) Finding 5 – AI/ML Issue Resolution Involves Changing Heterogeneous Artifacts:* Our study also suggests that issue resolution in AI/ML systems frequently extends beyond production code and involves heterogeneous artifacts, such as datasets, prompts, configuration files, checkpoints, and environment settings.

Among the 64 AI/ML issues, 28 issues (45%) required modifications outside production code. These changes include data cleaning and augmentation (3 issues), prompt engineering (5 issues), hyperparameter adjustments (3 issues), dependency and CUDA/library updates (7 issues), and modifications to CI, Docker, or runtime configurations (10 issues). In contrast, 83% of the non-AI/ML issues and 77% of the hybrid issues required only production code changes for resolution. For example, in AutoGPT's issue #2711 [38], developers resolved the issue by redesigning prompts and adjusting contextual inputs sent to the model rather than updating production code. Similarly, several issues required updating datasets or environment configurations to stabilize model behavior and improve reproducibility.

### C. Threats to Validity

Qualitative coding inherently involves subjective judgment. To mitigate this, two authors independently coded issues and associated artifacts, disagreements were resolved through consensus discussions moderated by a third author, and the codebook was iteratively refined throughout the study. Our findings are based on 100 closed issues from four open-source AI/ML projects. Although we selected projects from different domains and used random sampling to improve diversity, the findings may not generalize to other AI/ML systems. Third, several AI/ML-related activities and challenges appeared in only a subset of issues. Therefore, our findings should be interpreted as preliminary evidence of possible patterns rather than universally representative characteristics of AI/ML issue resolution.

## IV. Issue Resolution Tailored for AI/ML Systems

Building on the preliminary findings presented in Section III-B, we outline our vision for future issue resolution frameworks tailored for AI/ML systems. While traditional issue resolution frameworks [1], [22], [23] remain valuable, future AI/ML-tailored frameworks should extend beyond deterministic and code-centric assumptions to better support iterative experimentation, stochastic behavior, reproducibility, and coordination across heterogeneous AI/ML artifacts.

### A. What Future Frameworks Should Support?

The goal of adapting issue resolution frameworks for AI/ML systems is to better support the unique characteristics of such systems. AI/ML-tailored frameworks can help development teams design workflows that better fit the needs of their own AI/ML projects rather than relying on generic resolution processes. They can also improve reproducibility and traceability across datasets, production code, prompts, configurations, and models, while supporting more systematic experimentation and validation practices. Moreover, these frameworks can help practitioners, especially newcomers, get trained for and better understand how issue resolution unfolds in AI/ML projects.

Based on our findings, we identify five capabilities that future AI/ML issue resolution frameworks should support.

*1) Iterative and Cross-Stage Resolution Workflows:* Traditional issue resolution frameworks often assume a mostly sequential workflow across stages such as REPRODUCTION, ISSUE ANALYSIS, IMPLEMENTATION, and VERIFICATION [1], [3]. However, Finding 1 and Finding 2 suggest that AI/ML-specific activities such as MODEL PERFORMANCE MONITORING and PARAMETER TUNING AND TRAINING recur iteratively across multiple stages as developers repeatedly retrain models, reevaluate outputs, or modify datasets and configurations.

Future frameworks should therefore support repeated experimentation, reevaluation, and feedback loops rather than assuming strictly linear workflows. Such support is particularly important in AI/ML systems because developers may need to repeatedly validate fixes across multiple executions, datasets, and environments before gaining confidence in a solution.

*2) Reproducibility and Performance-aware Verification:* Finding 3 and Finding 4 show that reproduction and verification in AI/ML systems extend beyond deterministic pass/fail testing. Developers rely on repeated executions, statistical comparisons, qualitative inspection, and validation across different runtime environments before accepting a fix. Existing SE4ML research has similarly highlighted the challenges of reproducibility and nondeterministic behavior in AI/ML systems [6], [9].

Future frameworks should therefore provide explicit support for stochastic debugging, environment isolation, experiment replay, and performance-aware verification mechanisms. Such support can help developers systematically reason about variability introduced by datasets, random seeds, hardware, and dependency configurations, while reducing the current reliance on ad hoc experimentation and manual validation practices.

*3) Data-, Model-, and Artifact-aware Resolution:* As shown in Finding 5, resolving AI/ML issues requires modifications beyond production code, including updating datasets, prompts, hyperparameters, model checkpoints, and infrastructural settings. Future frameworks should treat these artifacts as first-class maintenance entities rather than auxiliary resources.

To support this, frameworks may need stronger provenance and traceability mechanisms that capture how models were trained, on which datasets, under which configurations, and within which environments. Existing tools such as MLflow [12], DVC [39], and Hugging Face [40] partially support experiment and artifact management, but these capabilities remain loosely connected to broader issue resolution workflows and collaborative debugging activities.

*4) Coordinating Heterogeneous Artifacts:* Finding 5 further suggests that AI/ML issue resolution often requires coordinated updates across interconnected artifacts such as production code, datasets, prompts, models, and infrastructure configurations. Changes in one artifact can influence downstream model behavior even when production code remains unchanged.

Future frameworks should therefore provide stronger traceability and provenance mechanisms that capture dependencies and relationships among heterogeneous artifacts [41]. With such support, developers can better understand how fixes propagate across the system and trace regressions or inconsistent outputs back to evolving datasets, prompts, or runtime environments.

*5) Human-AI Collaboration:* Given the growing adoption of LLMs and agentic systems in software engineering, future issue resolution frameworks will likely incorporate AI-assisted capabilities for issue triage, root-cause analysis, fix generation, and verification [5], [42]. However, our findings also suggest that AI/ML issue resolution frequently requires careful human judgment when interpreting nondeterministic outcomes, balancing predictive quality against resource usage, or deciding whether a fix is sufficiently reliable. As a result, future frameworks should support collaborative human-AI workflows rather than fully autonomous maintenance pipelines.

### B. Open Research Challenges and Opportunities

Several open research questions remain for realizing AI/ML-tailored issue resolution frameworks in practice.

*1) Understanding AI/ML Resolution Workflows:* Our study identified recurring AI/ML-related activities, challenges, and mitigation strategies, but these likely represent only a subset of real-world practices. Larger empirical studies are needed to examine how issue resolution varies across domains, project scales, and organizational settings. Future work should combine repository mining, developer surveys/interviews, and longitudinal studies [43] to better understand how practitioners resolve issues involving models, prompts, datasets, and deployment pipelines in practice. These findings can inform the development of AI/ML resolution frameworks.

*2) Validation Beyond Static Testing Procedures and Reproducibility Under Nondeterminism:* Our findings further suggest that traditional pass/fail testing procedures do not always align well with AI/ML systems, whose outputs may vary across runs and datasets. Future research should explore statistical evaluation methods, benchmark-aware verification, and human-in-the-loop evaluation techniques [44] that assess fixes across repeated executions and realistic deployment conditions. Recent work on evaluating AI systems and AI agents also highlights the growing importance of performance-aware and environment-aware evaluation workflows [45]. At the same time, future frameworks should support stochastic debugging [46], experiment replay, environment isolation, and probabilistic failure diagnosis [47] to address nondeterministic AI/ML behavior.

*3) Managing Heterogeneous Artifacts:* As shown in Finding 5, AI/ML issue resolution increasingly spans multiple interconnected artifacts. Future research should investigate automated and effective provenance, traceability, and dependency-management techniques that can help link changes across source code, datasets, prompts, models, and infrastructure configurations. Understanding how modifications propagate across these artifacts remains largely unexplored from an issue resolution perspective.

*4) AI-Assisted Maintenance for AI/ML Systems:* Current AI coding agents are primarily designed for traditional software workflows [42], but our findings suggest that AI/ML issue resolution often involves reasoning beyond production code alone. Future research should therefore explore AI-assisted maintenance agents capable of understanding heterogeneous artifacts, stochastic behavior, evaluation metrics, and experimentation workflows. Empirical studies are also needed to identify where existing agents fall short and how developers interact with AI-assisted maintenance workflows in AI/ML systems.

## ACKNOWLEDGEMENTS

This work was supported by the U.S. NSF grants CCF-2239107 and IIS-2533367. The views expressed are those of the authors and do not necessarily reflect those of the sponsors.